\documentclass{aa}
\input epsf
\newcommand{\sgr}{\mbox{V4641~Sgr }}

\begin{document}

\sloppypar

%
   \title{Super-Eddington outburst of V4641 Sgr}

   \author{M. Revnivtsev \inst{1,2}, M. Gilfanov \inst{1,2},
   E. Churazov \inst{1,2}, R. Sunyaev \inst{1,2}}

   \offprints{mikej@mpa-garching.mpg.de}

   \institute{Max-Planck-Institute f\"ur Astrophysik,
              Karl-Schwarzschild-Str. 1, D-85740 Garching bei M\"unchen,
              Germany,
	\and   
              Space Research Institute, Russian Academy of Sciences,
              Profsoyuznaya 84/32, 117810 Moscow, Russia
            }
  \date{}

        \authorrunning{Revnivtsev et al.}
        \titlerunning{Super-Eddington outburst of V4641~Sgr}
 
   \abstract{
X--ray transients provide unique opportunity to probe accretion
regimes of at a vastly different accretion rates. We analyze a
collection of the RXTE observations (Galactic Center scans, ASM
monitoring and a pointed observation) of enigmatic transient source
high mass X-ray binary V4641 Sgr and argue that they 
broadly support the hypothesis that 
giant September 1999 outburst was associated with an episode of
super-Eddington accretion onto the black hole. During the outburst an
extended optically thick envelope/outflow has been formed around the
source making the observational appearance of V4641 Sgr in many
aspects very similar to that of SS433. These results suggest that
objects like V4641 Sgr and SS433 indeed represent the class of 
objects accreting matter at a rate comparable or above Eddington 
value and the formation of an envelope/outflow is a generic
 characteristic of supercritical accretion. \\ 
 When the accretion rate decreased the envelope vanished and the source
short term variability and spectral properties started to resemble
those of other galactic black hole candidates accreting at a rate well
below the Eddington value. Interestingly that during this  
phase the source spectrum was very similar to the Cygnus X-1 spectrum
in the low state inspite of more than order of magnitude larger X--ray
luminosity.  
   }

   \maketitle

%

\section{Introduction}
The X-ray transient \sgr (SAX J1819.3-2525) 
was discovered by BeppoSAX and RXTE observatories in Feb 1999 
(\cite{sax_discovery},  
\cite{rxte_discovery}). During most of 1999  the source demonstrated only
weak X-ray activity (see e.g \cite{sax_paper}), until September 
1999 when series of very bright and short X-ray flares
occurred within less than $\sim 1.5-2$ days. During the brightest
flare, detected by the All  Sky Monitor aboard RXTE the source reached
the level of $\sim$12.2 Crab within a few hours
(\cite{asm_flare}). Subsequently the  X-ray flux from \sgr 
rapidly decreased and totally disappeared within less than $\sim 0.3$
day. On the decaying part of this giant outburst a pointed TOO RXTE 
observation of the source has been performed (see
e.g. \cite{markwardt_pca}).   

Analysis of optical images showed that the newly discovered
X-ray  transient can be associated with a variable star discovered by
Goranskij (\cite{gor78,gor90}, \cite{samus}). Previous unusual
activity of this star was observed in 1978  (Goranskij, 1978),
indicating that \sgr might be a recurrent transient with the
recurrence time of $\sim 20$ years. 

During the giant outburst in Sep., 1999 strong variability of \sgr  was
observed at all wavelengths  -- radio (\cite{hjellming99}),
optical (\cite{stubbings99}, \cite{kato99}), X-rays (1-12 keV) 
and hard X-rays (20--100 keV) (\cite{batse99}). The maximal observed 
optical brightness was at the level of $m_{\rm V}\sim8.8$
(\cite{stubbings99}). Spectroscopic observations in the optical band
performed during the outburst indicated a presence of a high velocity
wind (\cite{charles99}).

The VLA images, obtained on Sep.16.03, soon after the brightest 12.2
Crab flare (occurred on Sep.15.7, 1999), revealed an elongated extended
radio source with a  size of  $\sim 0.25\arcsec$
(e.g. \cite{hjellming00}). This led to a suggestion that \sgr might be
a new Galactic superluminal source. However, determination of the
jet velocity was complicated by quick decay of the radio  flux and the
absence of any direct observations of the apparent motion, in contrast
to the three previously known Galactic  superluminal sources
GRS1915+105, GRO J1655-40 and XTE J1748-288. 

Optical spectroscopy and photometry of the source performed during the
quiescence measured all  major parameters of the binary
system (\cite{orosz}, Table \ref{bin_pars}), including the mass of the
primary $M_{\rm primary}\sim$8.7--11.7$M_{\odot}$ and of the secondary $M_{\rm
secondary}\sim$5.5--8.1$M_{\odot}$, making \sgr a firm black hole
candidate in a binary system with a high mass companion.  
The distance to the source was constrained in the range of 7--12
kpc(\cite{orosz}, \cite{chaty}),  which exceeds significantly the
initial distance estimates of 0.5 kpc (e.g. \cite{hjellming00}). 
\cite{chaty} and \cite{orosz} infer slightly different distance 
estimates, but this difference do not influence strongly on our 
results. In our paper we will use the estimate obtained by \cite{orosz}.

Based primarily on the optical data: (i) significant brightening of the
source during the outburst, by $\Delta m_V\approx 5$, unusual for an 
HMXB transient and (ii) large peak optical flux, $m_{\rm V,
peak}\approx 8.5$, Revnivtsev et al. (2002) suggested that in Sept.1999 \sgr
might have had an episode of a super-Eddington
accretion. Super-Eddington accretion rate led to a formation of a 
massive optically thick and geometrically extended envelope/outflow
which enshrouded  the central black hole. The envelope, being optically
thick in the optical and X-ray bands absorbed/reprocessed the primary
emission of the central source and re-emitted bulk of the
accretion energy in the UV and EUV bands. Being the direct consequence
of near- or super-Eddington accretion onto the black hole, the
envelope vanishes during the subsequent evolution of the source when
the apparent luminosity drops well below the Eddington value.
In this paper we present the results of analysis of the data of Rossi
X-ray Timing Explorer data and demonstrate that the X-ray data
provides additional strong evidence in support of such  picture.

In Sects. 2 and 3 we describe results of analysis of the
RXTE data. In Sect. 4 we summarize the overall picture emerging from
the RXTE data and discuss the X-ray evidence of the extended optically
thick envelope surrounding the X-ray source during the maximum of its
activity in Sept.1999. 

\begin{table}
\caption{The parameters of the binary system \sgr (SAX J1819.3-2525). From \cite{orosz}.\label{bin_pars}}
\tabcolsep=0.4cm
\begin{tabular}{lr}
\multicolumn{2}{c}{Position}\\
\hline
R.A.=$18^h19^m21^s.64$&Dec.=$-25^{\circ}24\arcmin25\arcsec.6$ \\
$l$=6.77402&$b=$-4.78906 \\
\hline
\\
Parameter&Value\\
\hline
Orbital period, days&$2.81730\pm0.00001$\\
Mass function, ($M_{\odot}$) & $2.74\pm0.12$\\
Black hole mass, ($M_{\odot}$)& $9.61^{+2.08}_{-0.88}$\\
Secondary star mass, ($M_{\odot}$)& $6.53^{+1.6}_{-1.03}$\\
Total mass, ($M_{\odot}$)&$16.19^{+3.58}_{-1.94}$\\
Mass ratio &$1.50\pm0.13$\\
Orbital separation, ($R_{\odot}$)&$21.33^{+1.25}_{-1.02}$\\
Secondary star radius, ($R_{\odot}$)&$7.47^{+0.53}_{-0.47}$\\
Secondary star luminosity, ($L_{\odot}$)&$610^{+122}_{-104}$\\
Distance, kpc&7.4--12.3\\
\hline
\end{tabular}
\end{table}

\section{Observations and data analysis}

\begin{figure*}
\epsfxsize=17cm
\epsffile{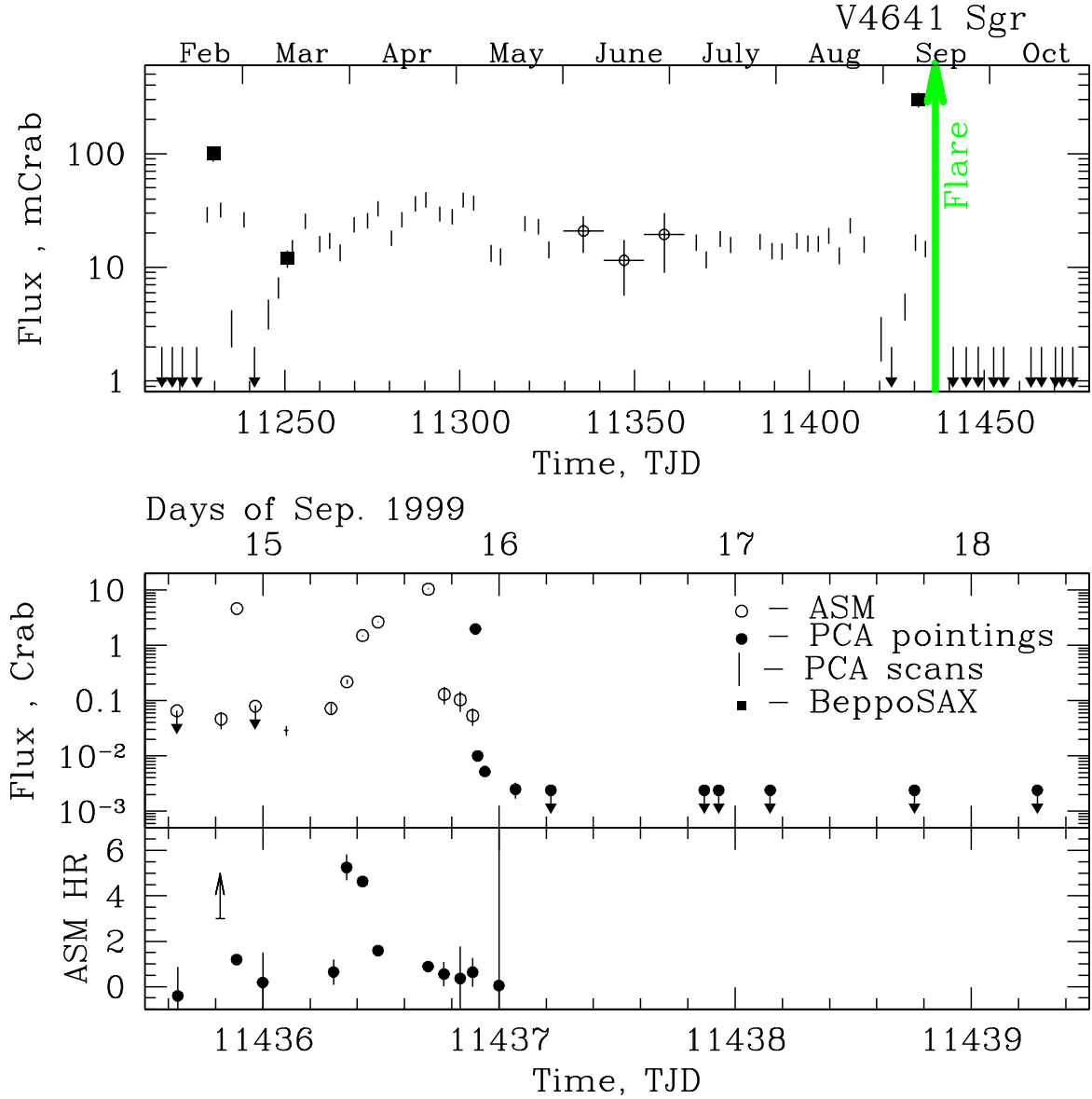}
\caption{The light curve of \sgr in 1999 according to observations of RXTE 
and BeppoSAX satellites. In the lower panels we present the source
ASM hardness ratio (5--12.2 keV)/(1.3--3.0 keV) during the period around the
bright X-ray  flares  ($\sim$Sep14-17, 1999). Filled circles represent   
the RXTE/PCA data of pointed observations, crosses - the RXTE/PCA scanning 
observations, open circles -- RXTE/ASM data, filled squares -- BeppoSAX data.
\label{lcurve}}
\end{figure*}

In our analysis we used publicly available RXTE data of \sgr.
This includes 11 pointed observations in Sep., 1999, 61 sets of the 
observations scanning over the Galactic bulge region, performed from Feb.5, 
1999 till Oct.24 1999, and three sets of scans over the Galactic bulge 
in Oct. 2000. 
In addition to this we used light curves of \sgr obtained by RXTE All 
Sky Monitor (ASM) in three energy channels (1.2--3--5--12.2 keV).
The RXTE/ASM lightcurves of \sgr  were provided by the RXTE/ASM team
 (http://xte.mit.edu/ASM\_lc.html).

The data reduction of PCA was performed with the help of standard procedures 
from FTOOLS/LHEASOFT 5.1 package. In order to avoid possible additional 
problems with the xenon edge at 4.8 keV (that became more pronounced with
 Epoch 4 response matrix) for our spectral analysis
we used only first layer data from all Proportional Counter Units (PCUs). 
To be convinced that our procedure allow us to perform the spectral analysis 
correctly we extracted the PCA spectrum of Crab Nebulae from the observation 
1999 Sep.13. Here we also used only the first layer data of all PCUs. The 
spectrum then was fitted by the conventional model: the power law with 
a photon index $\Gamma\sim 2.06$ with interstellar absorption 
($N_{\rm H}L$ was fixed at the value of 
4$\times10^{21}$ cm$^{-2}$). Residuals
 do not exceed the value of about 1\%. Therefore in 
subsequent analysis we used the value 1\% as a rough estimation of systematic 
uncertainties in the energy band 3--20 keV.
 In reality it appears that the uncertainties in the response matrix do 
not exceed 10\% at energies up to $\sim$40--50 keV therefore in some cases 
we used PCA data to illustrate approximate behavior of the source at these energies. For the
estimation of the PCA background we used VLE-based model when the source is 
sufficiently bright ($>$40 cnts/s/PCA) and L7\_240 model when the source is
weak. The nominal accuracy of the background estimation ($\sim$1--2\% of 
the background value, see e.g. 
http://lheawww.gsfc.nasa.gov/$\sim$keith/dasmith/Epoch4/ systematics\_l7\_240.htm)
 allows us to follow a source behavior down to a level of the order of
  0.1--0.5 mCrab. At such low level of the source activity the Galactic diffuse
emission can also become important. We estimated the contribution of the 
diffuse emission basing on the results of e.g. \cite{yk93}, \cite{y97} 
and found it small ($<$0.1-0.5 mCrab) at the position of the source
 ($l=6.774$, $b=-4.789$).

For the reduction of HEXTE data we used standard tasks of 
FTOOLS/LHEASOFT 5.1 package. For the HEXTE spectral analysis we used 
Archive mode (64 energy channels for total energy band 15-250 keV, 16 
sec time resolution). We analyzed Cluster~A and Cluster~B data separately 
and found that in each case their results are consistent with each other.
During fitting procedures we left normalizations of HEXTE clusters
 to be free parameters. Then, for plotting purposes the HEXTE different
 clusters points were averaged. Note here that absolute normalizations 
of HEXTE spectra on figures were adjusted to match the PCA spectra.

\section{RXTE results}

\subsection{Long term behavior of the source in 1999}

To follow the flux history of \sgr in 1999 we used publicly 
available data of RXTE/PCA scans over the Galactic Center region, that 
were performed
almost bi-weekly during the whole year. The statistical significance of the 
data and the accuracy of the background subtraction allow us to detect any 
source down to the level of approximately 1--2 mCrab (if the Galactic 
diffuse emission do not contribute much to the detected X-ray flux at 
the position of the source). The method of Galactic Center map construction 
and the extraction of the source flux is described in \cite{gravlens_paper}.

For the first time the \sgr was statistically significantly 
detected in PCA scan on Feb.18, 1999 
(\cite{rxte_discovery}) and since then it was seen in almost every scan. We 
present the obtained light curve of \sgr in Fig. \ref{lcurve}. The results of
our analysis (61 data points) are in agreement with the 7 data points 
reported in \cite{rxte_discovery} and \cite{markwardt_pca}.

After almost 6 months of moderate X-ray activity,  during the
 first half of Sep.1999 \sgr demonstrated several X-ray outbursts.
The first, the weak one, was detected by BeppoSAX and RXTE/ASM on 
Sep.10.1, 1999. The source X-ray flux  reached $\sim$300 mCrab 
(\cite{sax_paper}). On Sep.14--15, 1999  three more powerful flares 
were detected. The segment of the light curve of \sgr around 
Sep.14--15, 1999 is presented in Fig. \ref{lcurve}(lower panel).
It is seen that the source rose up to 4, 12 and 2 Crabs on Sep.14.9, 
Sep.15.7 and Sep.15.9 respectively. The observed large changes in X-ray flux
(factor of 10 at least) occurred on the time scales of one--two hours.
The X-ray luminosities of the source in the energy band 1--12 keV
during these outbursts could be
estimated to be $\sim1\cdot10^{39}$ ergs/s, $\sim3.4\cdot10^{39}$ and $\sim7\cdot10^{38}$ ergs/s respectively (with adopted distance to \sgr $\sim$9.5
kpc, \cite{orosz}).

The last X-ray flare finished on $\sim$Sep.15.95, 1999 by the rapid drop 
in X-ray flux from the level of $\sim$100 mCrabs by a factor of 10 and 
after this the source returned to the quiescent state - $\la$1 mCrab.
Note, that at such low flux levels some contamination from
the Galactic diffuse emission is possible. Our estimates of the  
contribution of the Galactic diffuse emission at the position of \sgr 
showed that it is small, but not negligible. As a very conservative
 estimation of the source flux the detected 1 mCrab flux at the 
position of \sgr should be treated as the upper limit.

\subsection{Orbital modualtion of the X-ray flux}

\begin{figure}
\epsfxsize=8.5cm
\epsffile[50 170 570 540]{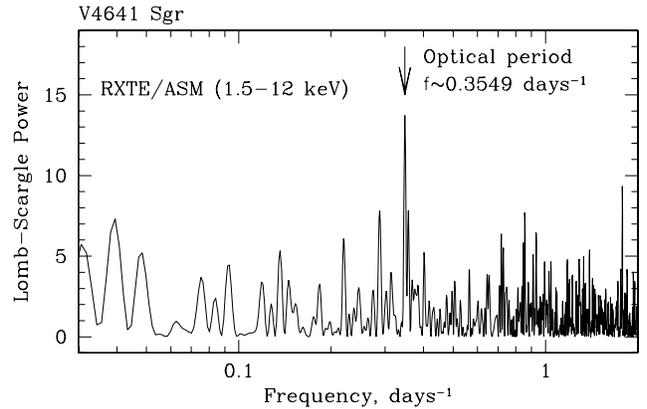}
\caption{The Lomb-Scargle periodogram of the lightcurve of \sgr, during 
period of moderate source activity, Mar.13-Aug.20, 1999. The value of the optical period is marked by an arrow.\label{lomb}}
\end{figure}

In spite of the fact that the significance of the source detection in
the RXTE/ASM data is lower than in the RXTE/PCA  
data (during  one day of observations), ASM points have advantage from
the point of view of their  quasi-uniform coverage over the period
of Feb.-Sep., 1999. This helps us to 
search for the source period in X-ray data. Periodicities of \sgr were
sought by means of the Lomb-Scargle periodograms (\cite{lomb},
 \cite{scargle}, \cite{press}). In Fig. \ref{lomb} we present the Lomb-Scarge
periodogram obtained for \sgr lightcurve taken in the period of steady state
activity $\sim$Mar.13-- Aug.20, 1999 ($\sim$TJD11250-11410)

It is seen that almost at the position of detected optical period 
$P=2.8173$ days(\cite{orosz}) a peak at the X-ray Lomb-Scarge periodogram is 
present. This peak corresponds to the period $P_{\rm x}=2.84\pm0.03$ days. 
The uncertainty of the period value was estimated by a Monte-Carlo 
bootstrap method, assuming the gaussian distribution of values of ASM 
lightcurve points.
The false-alarm probability of this detection (taking into account the 
number of trial periods) is slightly less than $10^{-3}$,
if we assume the exponential distribution of Lomb-Scargle power values.
It is not very high significance to rely on X-ray data alone. 
However, the marginally detected X-ray periodicity has the value of the 
period that coincides within 1-$\sigma$ errors with the firmly detected optical one $P=2.8173$ days. This strongly supports the detection of the binary period in X-rays.

\begin{figure}[htb]
\epsfxsize=8.5cm
\epsffile[50 170 570 540]{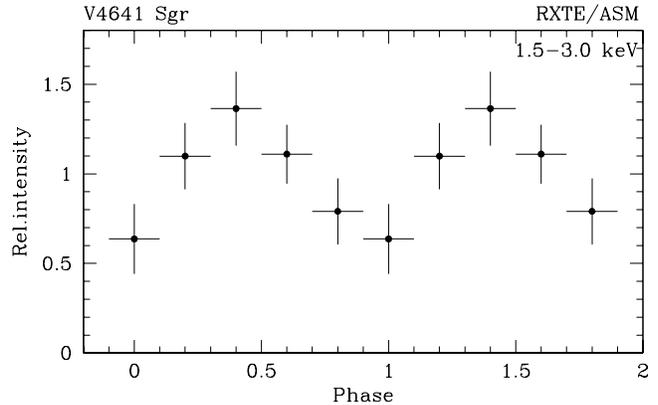}
\caption{The X-ray lightcurve of \sgr in the energy band 1.5-3.0 keV folded
 with the optical photometric period $P=2.8173$ days. The reference 
time $T_0$ taken from the paper of Orosz et al.(2001). The phase 0 corresponds to time when the normal star is located between the black hole and observer.\label{folded}}
\end{figure}

We have folded the lightcurve of \sgr with the measured optical
period of the system in order to search for the orbital dependence of the 
X-ray flux. The obtained orbital profile of the X-ray flux(1.5-3.0 keV, 
lowest ASM energy channel) is presented in 
Fig. \ref{folded}. In order to make the comparison of optical and X-ray 
folded lightcurves easier we have used the same reference time $T_0$ as
 Orosz et al.(2001) in their Fig. 3. It is seen that the folded X-ray 
lightcurve demonstrates peak, when the black hole is located between the 
observer and the optical star, and the minimum - when the star is located 
between the observer and the black hole. We also detected strong dependence
 of the amplitude of orbital modulations of X-ray flux on the photon energies:
for the lowest energies, 1.5-3.0 keV, the amplitude of the sinusoidal 
variations is $36\pm8$\%, in the energy band 3-5 keV -- $25\pm5$\%, 
and in the energy band 5-12 keV the modulations is undetectable with an
upper limit $<15$\%(2$\sigma$). It should be noted, that the observed
energy dependence of the X-ray orbital modulations and the position of the
X-ray minimum on this modulation suggest that the detected X-ray variations
could be caused by an absorption in the line of sight near the optical star.
This suggests that the inclination of the system is close to
65--70$^{\circ}$, in agreement with the optical data (\cite{orosz}).

\subsection{The spectral evolution of the source}

\subsubsection{Period of ``quiescent'' activity in Feb.-Sep.,1999}

\begin{figure}[htb]
\epsfxsize=8.5cm
\epsffile[30 180 570 650]{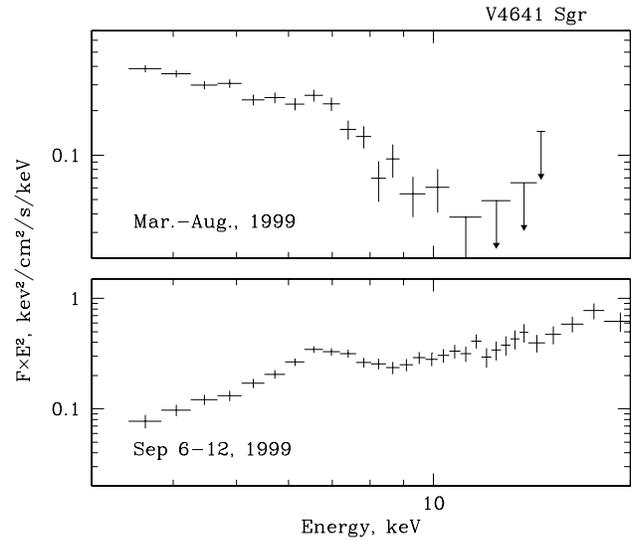}
\caption{The spectrumn of \sgr averaged over period Mar.- beginning of Sep. 1999.
 The lower spectrum represents the set of spectra obtained during 
the period Sep.6-Sep.12, 1999. \label{spec_5may}}
\end{figure}

The scan observations give us an important opportunity to 
follow the spectral shape of the source during a year. 
Unfortunately the acceptable ``on-source'' 
exposure of a scan observation is 
only of the order of 10--20 secs. Therefore the statistics in the
obtained spectra is quite poor. During the period Feb.-Sep. 1999
the source was detected with quite soft spectrum -- which 
could be roughly described by the model of bremsstrahlung
 emission with temperatures $kT\sim$2--3 keV or by multicolor disk model
(\cite{ss73}) with the inner disk temperature $kT\sim1.5$ keV. 
The typical spectrum of \sgr at that time is presented in 
Fig. \ref{spec_5may}(upper panel).  

The spectrum of \sgr
obtained by BeppoSAX observatory on Mar.13, 1999 (see \cite{sax_paper})
 has the statistically 
significant emission line at the energy $\sim7$ keV with 
equivalent width $\sim$270 eV. Therefore we also searched 
for the emission line in the PCA scan data. Unfortunately, due to poor
 statistics, the emission line could not be detected in a single
 PCA scan observation with an upper limit on its equivalent width 
$EW<$0.7-1.0 keV. 
However, a fit to spectrum of \sgr, averaged over the period Feb.18--Sep.02,
1999 gives an emission line at the energy 
$E_{\rm line}=6.60\pm0.08$ keV with the equivalent width $EW=360\pm90$ eV.

\begin{figure}[htb]
\epsfxsize=8.5cm
\epsffile[10 180 570 520]{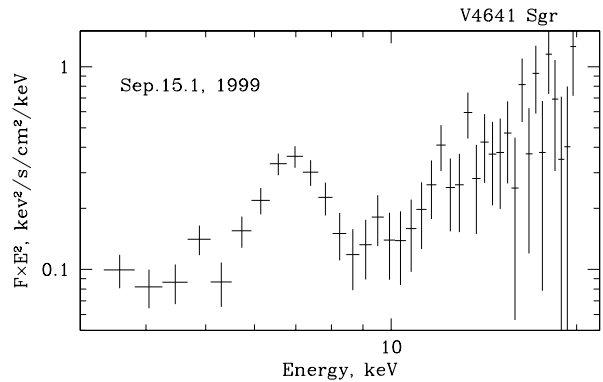}
\caption{The spectrum of \sgr on Sep 15.1, 1999, between the two
bright X-ray flares (see Fig. \ref{lcurve}). \label{spec_15sep}} 
\end{figure}

The X-ray spectrum of the source strongly changed in the beginning of 
Sep 1999 after the dip in the light curve (see Fig. \ref{lcurve}). The photon 
spectral index hardens - to $\alpha\sim1$  - and the emission line
became stronger - $EW\sim$1--2 keV (with typical uncertainty 
$\sigma_{\rm EW}\sim$200-300 eV). The spectrum averaged over Sep.6-12, 1999
is presented in Fig. \ref{spec_5may}(lower panel). The spectrum resembles the 
one obtained by BeppoSAX on Sep.10, 1999 (\cite{sax_paper}). Note, that
approximately simultaneously with the dramatic changes in the X-ray
spectral properties of the source the increase of the source optical activity
was detected (\cite{kato99}).

\subsubsection{Period of flaring activity (Sep.14.8-15.9, 1999)}

During the outburst activity on Sep.14-15, 1999, the source demonstrated 
at least three powerful X-ray flares (Fig. \ref{lcurve}), two of which
were detected by the ASM instrument. 
ASM data indicated that during the flares the spectral hardness of the
source was generally anticorrelated with the X-ray flux.

Between the two ASM flares, on $\sim$Sep.14.9,  the source
flux dropped by a factor of 20 at least and became undetectable by the 
ASM. Fortunately, at this time, on Sep.15.1, 1999 a PCA scan
observation was performed, allowing us to investigate  
the spectrum of \sgr between the two flares. 
During this observation a  remarkable spectrum was obtained
(Fig. \ref{spec_15sep}). The spectrum is dominated by the emission line
at $E_{\rm line}=6.63\pm0.08$ keV with enormous equivalent width of 
$EW=2.4\pm0.3$ keV. Remarkably, this spectrum is very similar to an
X-ray spectrum of SS433 (e.g. \cite{margon84}).

\subsubsection{The pointed RXTE observation (Sep. 15.89-15.95, 1999)}

The third of the detected bright flares occurred during the decaying
part of the outburst and was missed by the ASM because of $\sim$1.5
hour gap between the ASM points. 
Coincidentally, it occurred during the pointed observation of RXTE.
As it was previously mentioned by \cite{markwardt_pca} the 
spectrum of the source at the peak of the X-ray light curve during this
RXTE observation was quite hard and resembled the typical spectra of
the black holes in the low/hard spectral state with the 3--50 keV
photon index $\alpha <$2.0, the cutoff at the 
energies 100-200 keV, and pronounced fluorescent Fe line at 6.4 keV.

\begin{figure}
\epsfxsize=8.5cm
\epsffile[20 180 400 720]{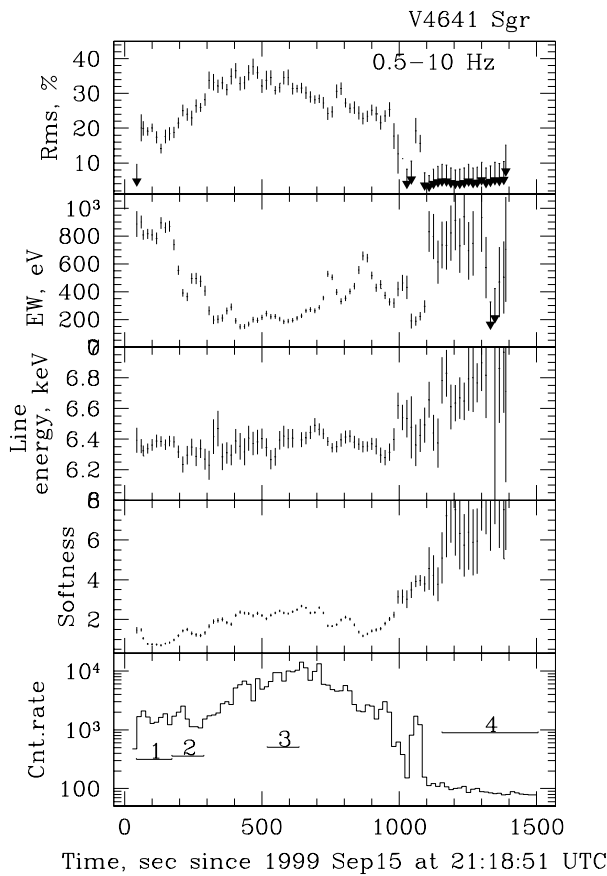}
\caption{The spectral evolution of \sgr during the  first 1500 sec of the 
RXTE pointed observation. The solid lines in the lower panel represented the 
intervals used for the accumulation of the broad band spectra of \sgr (see. Fig. \ref{spectra})
\label{parameters}}
\end{figure}

\begin{figure}
\epsfxsize=8.5cm
\epsffile[0 170 570 720]{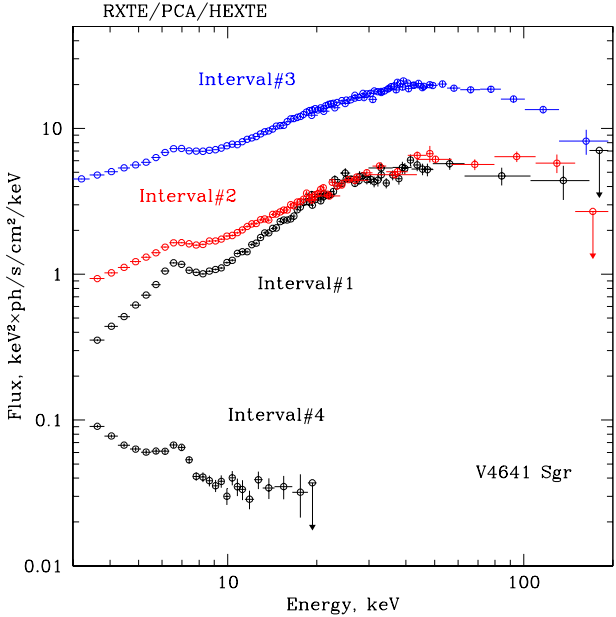}
\caption{The spectra of \sgr, accumulated over time intervals, shown in Fig. \ref{parameters}. The change in the
 hardness and in the strength of the fluorescent line is clearly 
seen.\label{spectra}}
\end{figure}

However, time resolved spectral analysis of the RXTE/PCA data revealed 
significant quantitative and qualitative evolution of the source
spectrum during $\approx 1500$ sec of the pointed RXTE observation.
In Fig. \ref{parameters} we present the light curve of the
source with 16 sec time resolution, softness ratio (3--5 keV to 15--20 keV),  
the centroid energy of the Gaussian line and its equivalent width
as a function of time. The position of the line and its 
equivalent width were determined using a simple power law + Gaussian line 
approximation of the spectrum in the 3--12 keV energy band. 
In the uppermost panel we also show behavior of the integrated  
fractional rms (in percents) of the source flux variations (3--20 keV
energy band, 0.5-10 Hz frequency range). 
In order to illustrate the spectral evolution of \sgr
we show in  Fig. \ref{spectra} the broad band spectra accumulated
during four intervals, marked in Fig. \ref{parameters} (lower panel) by
the horizontal lines.

As is apparent from Figs.\ref{parameters} and \ref{spectra} the RXTE
observation can be divided into two parts with qualitatively different
spectral properties with the boundary at $t\sim 1100$ sec (in the
units of Fig. \ref{parameters}) corresponding to the final drop of the
X-ray flux and change of the iron line energy from $\approx 6.4$ keV
to $\approx 6.6-6.8$ keV.

\begin{figure}
\epsfxsize=8.5cm
\epsffile[52 180 570 550]{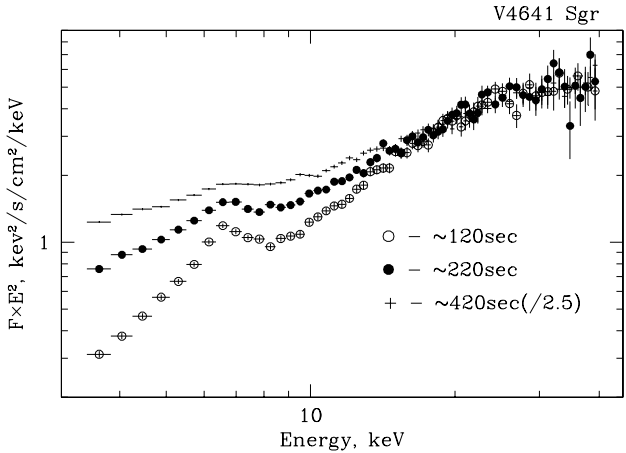}
\caption{Three spectra of \sgr obtained at different times. 
Open circles represent the spectrum obtained at $\sim$120-th sec 
(in time units used if Fig. \ref{parameters}), filled circles - at
 $\sim$220-th sec, and crosses - at $\sim$420-th sec, scaled down 
by factor of 2.5 to match the other two at energies higher than $\sim$20 keV.
All spectra were accumulated over 16 sec time intervals.
\label{set_of_spec}}
\end{figure}

During the first part the source had a strong emission line centered 
at $\sim$6.4 keV with the equivalent width varying between 200 and 900
eV, sufficiently hard spectrum extending to the hard X-ray energies
and significant aperiodic variability with fractional rms $\sim 15-40
\%$. In order to qualitatively  illustrate the character of the
spectral evolution, we show in Fig. \ref{set_of_spec} three spectra,
accumulated during individual 16 sec intervals, 
corresponding to significantly different values of the line equivalent
width and fractional rms. 

The Figs.\ref{parameters} and \ref{set_of_spec} demonstrate that the
spectral evolution can be qualitatively understood as a result of
absorption/reprocession in the extended medium with varying absorption
column density. Indeed, assuming that the primary spectrum does not
change significantly, decrease of the absorption column density
would lead to the apparent softening of the outgoing spectrum and
decrease of the equivalent width of the fluorescent iron
line. If the absorbing/reprocessing medium has a
significant spatial extend with the light crossing time of $\sim
10-50$ sec. the variations of the primary emission would be smeared
out in the reprocessed emission, the effect depending on the fraction
of the scattered/reprocessed emission in the outgoing radiation. Thus,
decrease of the absorption column density would lead to increase of
the apparent fractional rms. It should be noted, that 
absorption by the neutral medium with solar element abundances does
not adequately explain the observed spectra -- certain ionization of
the absorbing gas is required by the data. The maximal  $N_{\rm H}L$
value should be of the order of $\sim$few$\times10^{23}$cm$^{-2}$.

We have not found any strong soft component in the spectrum during
this part of the observation. However, this issue is rather
complicated taking into account the absence of the spectral data at
energies lower than 3 keV.

After the final drop of the X-ray flux at $t\sim 1100$ sec a
significant softening of the spectrum occurred and the line shifted
from $\approx 6.4$ keV to $\approx 6.6-6.8$ keV indicating a
significant change of the emission regime.

\begin{figure}
\epsfxsize=8.5cm
\epsffile[52 350 570 700]{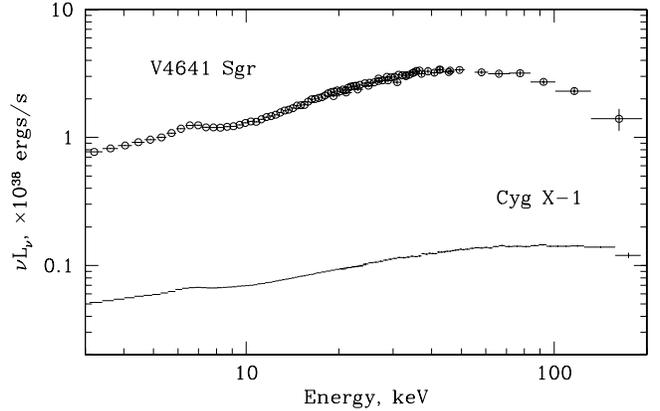}
\caption{Comparison of the spectra of \sgr, accumulated during the peak
of the observed light curve (Fig. \ref{parameters} and Cyg X-1 in the
hard state.  The distance to Cyg X-1 was assumed to equal to 2.5 kpc.  
\label{comparison_with_cygx1}}
\end{figure}

At the maximum of X-ray lightcurve, when the absorption 
was presumably weak the source had ordinary hard spectrum,
typical for black holes in the low spectral state 
(Fig. \ref{spectra} and \ref{comparison_with_cygx1}). Remarkably,
\sgr have demonstrated  this type of spectrum while it's luminosity exceeded by
$\sim$10 times that of Cyg X-1 in the hard and, probably by $\sim$3-5
times in the soft spectral states.  Let us mention that masses of black 
holes in both systems are comparable. Therefore the difference in 
luminosity might give an information about the accretion rate.

\subsection{Short-term variability}

The only data suitable to study short term variability of \sgr during
the period of flaring activity in Sept.1999 are that of the pointed
RXTE observation discussed in the previous subsection.
The brief description of the source variability during this
observation can be found in \cite{wvdk}.
During the first $\sim$1100 sec of the observation the \sgr was 
found to be strongly variable (rms amplitude $\sim$50\%) 
with a power law power spectrum in the $10^{-2}-5$
Hz frequency range. At the frequency of $\sim5$ Hz the power 
spectrum changed from $P\sim f^{-1}$ to $P\sim f^{-2}$. It is
interesting to note that in spite of similarity of the spectral
properties of \sgr with the spectral properties of Cyg X-1 in the hard
state  the power spectrum of \sgr flux variability is more similar to
that of Cyg X-1 in the soft state.  

We discuss below change of the aperiodic variability properties with
time, photon energy dependence of the fractional rms and time delay of
the reflected emission.

As was shown in the previous subsection, during the RXTE
observation the source demonstrated a strong spectral evolution 
accompanied with significant change of the fractional rms
(Fig. \ref{parameters}). Similar to the spectral properties,
the variability level changed significantly 
after the rapid drop of the X-ray flux at $t\sim$1100-th
sec (Fig. \ref{parameters}). The fractional rms droped from $\sim
30-40\%$  to the level, undetectable with PCA, with the $2\sigma$
upper limit of $\sim$1--2\% in the $5\cdot10^{-3}-10$  Hz frequency
band.

\begin{figure}
\epsfxsize=6.5cm
\epsffile[0 140 470 710]{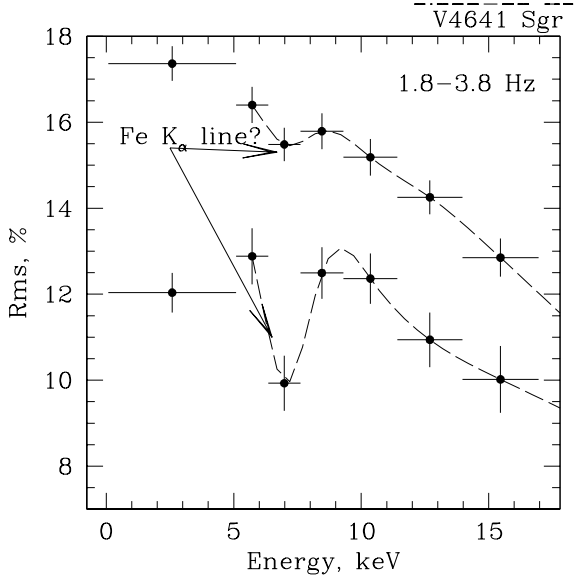}
\caption{The dependence of the rms amplitude of X-ray variability of \sgr on
 the photon energy during the episode of low equivalent width of the Fe line
during two time intervals -- $<$300-th sec (lower points) and 
$\sim$350--700-th sec (upper points) in the units of Fig. \ref{parameters}. 
The dashed line are the splines for the clarity. Dips at the place of Fe line mean that the line flux is less variable than the continuum flux. \label{rms_en}}
\end{figure}

A notable feature of the energy dependence of the fractional rms is
the decrease near the energy of the Fe $K_{\alpha}$ line. 
Note that some indications on such behavior could be noticed in Fig. 3
of \cite{wvdk}. However the authors concentrated on the properties
averaged the entire outburst.
In Fig. \ref{rms_en} we present the dependence of rms amplitude of the
source variability calculated for 
two different periods - intervals $<$300-th second in the time units
of Fig. \ref{parameters} (high EW of the line) and $\sim$400--700-th
sec  (low EW of the line). One can see that there exist a definite dip
approximately  at the position of the Fe 6.4 keV fluorescent
line. Moreover, it is seen that  
the stronger the line in the source's energy spectrum,
 the stronger the dip at the rms-energy dependence.
The presented rms-energy dependence clearly demonstrate that the 
fluorescent Fe line is less variable than the continuum. 
The simplest interpretation of this fact could be the smearing of the 
reprocessed emission (in particular the flux in the Fe fluorescent line) 
because of the finite light crossing time of the reprocessing medium 
(see e.g. discussion in \cite{mikej_feline}, or \cite{marat00}).

\begin{figure}
\epsfxsize=8.5cm
\epsffile[32 170 570 550]{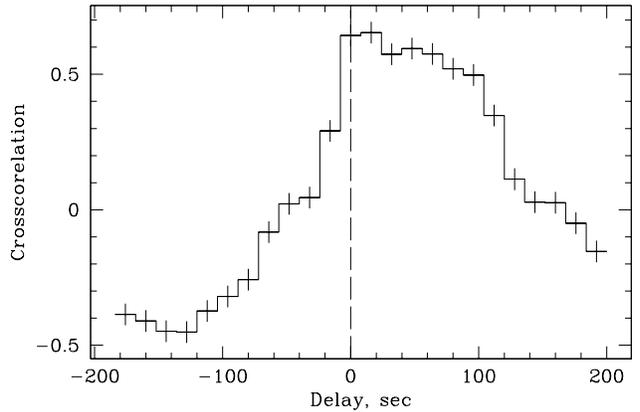}
\caption{The crosscorrelation between the continuum X-ray flux from
 \sgr and the flux in the Fe fluorescent line, calculated over period
$\sim$350--700 secs in the time units of Fig. \ref{parameters}. The
error bars were estimated  using Monte Carlo bootstrap method. The
positive values of delay correspond to the Fe line flux being delayed
with respect to the continuum.\label{crosscor}} 
\end{figure}

The reprocessing in medium of large light crossing time can also
result in the time delay  between the direct continuum emission and
the photons of the reprocessed  spectrum, in particular -- the Fe line
photons. In order to check this hypothesis 
we have crosscorrelate the continuum flux of \sgr and the flux in the 
fluorescent Fe line. Flux in the Fe line was taken from the spectral
approximation used for Fig. \ref{parameters}.
In order to avoid possible contamination by long term trend in the
parameters we used only data during the period with relatively stable
value of  equivalent width of the fluorescent line -- from
$\sim$350-th to $\sim$700-th seconds (in the time units of
Fig. \ref{parameters}). The obtained crosscorrelation is presented in
Fig. \ref{crosscor}. The  
time interval of the acceptable data is  rather short and we could not 
calculate the uncertainties on the crosscorrelation function directly
from the data. Therefore we estimated the error bars using Monte Carlo
bootstrap method.  From Fig. \ref{crosscor} one can see that the
crosscorelation function is strongly asymmetric with respect to the
zero delay, implying that the flux in the Fe fluorescent line is
delayed with respect to the continuum flux. The  approximate time of
the delay is $\tau\sim$50 sec.

\section{Discussion}

\subsection{\sgr -- Summary}

\begin{figure}
\epsfxsize=8.5cm
\epsffile[32 170 550 550]{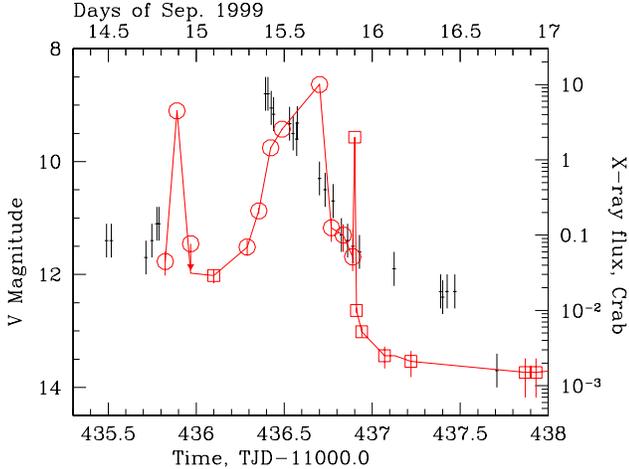}
\caption{The light curves of \sgr in optical (VSNET data, http://www.kusastro.kyoto-u.ac.jp/vsnet/Xray/gmsgr.html) and X-ray energy bands.
The RXTE/ASM points are shown by the open circles, the RXTE/PCA points by open squares. \label{optics}}
\end{figure}

Based on the available RXTE data of the Galactic Center scans, ASM
monitoring and RXTE pointed observation the overall picture of the
\sgr evolution can be summarized as follows:  

\begin{enumerate}

\item During approximately a half of a year after its discovery \sgr
remained in the ``quiet'' state (Fig. \ref{lcurve}) with X-ray 
luminosity at the level of $L_X\sim 10^{36}$ erg/sec (flux $\sim
10-20$ mCrab) and a soft spectrum (Fig. \ref{spec_5may}).

\item A ``landmark'' in \sgr evolution was a dip which
started on Sept.2 and lasted for $\sim 10$  days (Fig. \ref{lcurve}). 
The source luminosity droped by a factor of $\ga 10$. After the dip
the source spectrum changed dramatically (Fig. \ref{spec_15sep}) and
the optical emission increased by $\Delta m_V\sim$2 (\cite{kato99}). 

Interestingly, the apparent size of the radio jet can be reconciled
with the constrains on the binary system inclination angle only
assuming that the ejection occurred before Sept.6. It seems
plausible to associate the jet ejection with the X-ray dip. 

\item On Sept.14.9 a period of flaring activity started 
(Fig. \ref{lcurve}, lower panel, Fig. \ref{optics}).  Although, due to
fastness of the events, the data are rather sparse, comparison of the
optical and X-ray light curves (Fig. \ref{optics}) suggests
sufficiently smooth evolution of the optical light, probably
reflecting change of the bolometric luminosity and/or mass accretion
rate. The X-ray flux, on the contrary, changed in a rather irregular
fashion. Three bright X-ray flares were detected during rising and
decaying parts of the optical light curve, with a deep minimum at the
epoch of the maximal optical/bolometric luminosity.
The source spectrum (Fig. \ref{spec_15sep}) during the X-ray
minimum had a strong emission line at $\approx 6.6$ keV with
equivalent width of $\approx 2.4$ keV, indicating dominant
contribution of the reflected/reprocessed emission.
During the brightest flare the source reached
flux level of 12.2 Crab corresponding to the 1-10 keV luminosity of
$\sim (3-4)\cdot 10^{39}$ erg/sec or exceeded Eddington luminosity 
for a 10 $M_{\odot}$ black hole.

\item The final phase of the third flare was studied by collimated
instruments aboard RXTE. The spectral evolution during the flare can
be understood in terms of 
absorption by ionized medium of varying column density. Strong
fluorescent line of iron present in the spectra had smaller fractional
rms than the continuum flux and was delayed with respect to the
continuum by $\sim 25-50$ sec. 

\item In the period of low absorption the source demonstrated a
hard spectrum typical for black holes in the low spectral state 
(Fig. \ref{comparison_with_cygx1}). Its luminosity, however, exceeded
by a factor of $\sim 10$  and $\sim 2-3$ respectively the Cyg X-1
luminosity in the hard and the soft spectral states.

\item An orbital modulation of the X-ray flux during the period of
quiescence was detected with fractional amplitude of $\sim 30\%$. 
The phase and energy dependence of the orbital modulation is
consistent with obscuration by the matter in the vicinity of the
optical companion, in which case the inclination of the binary system
should be close to $\sim 65-70\degr$

\end{enumerate}

\subsection{Super-Eddington accretion and optically thick envelope}

Earlier, we have suggested (\cite{v4641_opt}), that the optical data collected
during the period of flaring activity of \sgr can be naturally
understood assuming formation of an optically thick warm ($T\sim 10^5$
K) envelope/outflow enshrouding the central source. The envelope
is a direct consequence of significantly super-Eddington accretion and
disappears when the mass accretion rate decreases below the Eddington
value. Such an envelope, being
optically thick at the optical wavelengths due to free-free processes
and in the X-ray band due to absorption by the metals and Compton
scattering, absorbs and re-emits bulk of the central source
luminosity. In such a picture, smooth behavior of the optical light
reflects evolution of the bolometric 
luminosity and, possibly, of the mass accretion rate. The bright X-ray
flares occurred on the rising and decaying part of the
optical light curve are a result of the changes in geometry and/or
optical depth of the envelope. 

At the peak of the optical light
(which likely corresponded to the peak of the bolometric
luminosity and the maximum in the optical depth of the envelope) the
central source was almost completely obscured, which caused deep
minimum in the X-ray flux. The weak X-ray emission observed in the
scanning observation could be a result of reprocession of the primary
X-rays by surrounding warm Compton thick gas. 
This is in good qualitative agreement with the
actually observed spectrum (Fig. \ref{spec_15sep}) which is extremely hard and has
very strong fluorescent line of iron with equivalent width of $\approx
2.4$ keV. The line centroid energy, $E\sim6.7$ keV requires significant
ionization of iron. An alternative explanation of the observed large
equivalent width of the line could be thermal emission from optically
thin gas in the base of the jet, similarly to interpretation of
the spectra of SS433. The picture could be clarified if the hard X-ray 
data was available. 

The only broad band spectroscopical data available is that of the RXTE
pointed observation performed in the end of the outburst.
Coincidentally, it caught the tail of a short flare occurred during
decaying part of the outburst. The spectral evolution observed by RXTE
can be understood as an effect of absorption by ionized gas with
the column density decreasing with time. This picture is further
supported by the study of variability of the iron K$_{\alpha}$
emission. We found that fractional  rms of the iron K$_{\alpha}$
flux is smaller than that of the surrounding continuum (Fig. \ref{rms_en}). This
result can be easily understood as smearing of the variations on the
time scales shorter than the light crossing time of the reprocessing 
media. In addition, the finite light crossing time of the reprocessing
media should cause time delay of the  reprocessed emission
with respect to the primary flux -- in a good agreement with the
observed behavior (Fig. \ref{crosscor}). The amplitude of the observed time delay,
$\tau\sim 50$ sec, corresponds to the linear size of $\sim 10^{12}$
cm, i.e. is of the order of the Roche lobe of the binary system, as
should be expected.  

\section{Conclusions}

\begin{enumerate}
\item We argue that X--ray data are broadly consistent with the assumption
that 1999 September outburst of V4641 was caused by the episode of 
super-Eddington accretion onto a black hole accompanied by a formation of
an extended envelope. The properties of the source during this outburst 
are similar to those of SS433 suggesting that formation of the 
envelope/outflow is a generic characteristic of the supercritical 
accretion as predicted by many theoretical models. 

\item It is interesting however that hard spectrum, resembling the spectrum 
of Cygnus X-1 during low state, was observed from V4641 at the
luminosity level more than order of magnitude larger than in Cygnus X-1.

\item Using ASM data during the extended period of relatively quiescent 
state of the source prior to the giant outburst we marginally 
detected periodicity
in the 1.5-12 keV energy band, consistent with the optically determined
orbital period.

\item Almost two weeks prior to outburst dramatic evolution of spectrum is
observed, associated with the significant decrease of the source X-ray
flux. If the subsequently observed extended radio source is due to
ejection happened at this time (and not during the major outburst), 
then expansion velocity would be at the level of 0.7-0.8 c -- similar to
the velocity observed in several galactic jet sources.
\end{enumerate}

\begin{acknowledgements}
This research has made use of data obtained through the High Energy
Astrophysics Science Archive Research Center Online Service, provided 
by the NASA/Goddard Space Flight Center. 
\end{acknowledgements}

\end{document}